\documentclass[aps,prl,twocolumn,groupedaddress,superscriptaddress,
			   amsfonts,amssymb,amsmath,citeautoscript,a4paper
			   ]{revtex4-1}
\pdfoutput=1 

\usepackage{microtype} 
\usepackage{xspace} 

\usepackage{txfonts}  
\usepackage{txfontsb} 

\usepackage{bm} 
\usepackage{moresize} 

\usepackage{xcolor}
\usepackage{graphicx}
\usepackage{tikz}
\usetikzlibrary{decorations.text}

\usepackage[]{booktabs}
\usepackage{array}
\usepackage{layouts}
\definecolor{mygray}{rgb}{.15 .15 .15}

\usepackage{hyperref}
\hypersetup{
    colorlinks,
    linkcolor={blue!75!black!80!yellow},
    citecolor={blue!75!black!80!yellow},
    urlcolor={blue!75!black!80!yellow}
}

\usepackage[centering,hmargin=16mm,tmargin=30mm,bmargin=26mm]{geometry}

\newcommand{\rv}{\mathbf{r}}
\newcommand{\nv}{\hat{\mathbf{n}}}

\newcommand{\vf}{v_{\text{\textsc{f}}}}

\newcommand{\e}{\mathrm{e}}

\newcommand{\dde}{\,\mathrm{d}}
\newcommand{\appropto}{\mathrel{\vcenter{
  \offinterlineskip\halign{\hfil$##$\cr
    \propto\cr\noalign{\kern.2pt}\sim\cr\noalign{\kern-2.5pt}}}}}
\newcommand{\spar}{{\scriptscriptstyle\parallel}}
\newcommand{\sperp}{{\scriptscriptstyle\perp}}

\newcommand{\dperp}{d_{\!\sperp}}
\newcommand{\dpar}{d_{\spar}}

\newcommand{\iu}{\mathrm{i}}

\makeatletter
\newcommand{\raisemath}[1]{\mathpalette{\raisem@th{#1}}}
\newcommand{\raisem@th}[3]{\raisebox{#1}{$#2#3$}}
\makeatother

\newcommand{\ie}{i.e.\@\xspace} 
\newcommand{\cf}{cf.\@\xspace}
\newcommand{\eg}{e.g.\@\xspace}

\makeatletter
\renewcommand{\fnum@figure}{\figurename~\thefigure\ (color online)}
\makeatother

\begin{document}
\title{Quantum corrections in nanoplasmonics: shape, scale, and material}

\author{Thomas~Christensen}
\email{tchr@mit.edu}
\affiliation{Department of Physics, Massachusetts Institute of Technology, Cambridge, Massachusetts, USA}
\author{Wei~Yan}
\email{wyanzju@gmail.com}
\affiliation{Institut d'Optique d'Aquitaine, Universit\'e Bordeaux, CNRS, 33405 Talence, France}
\author{Antti-Pekka Jauho}
\affiliation{Center for Nanostructured Graphene, Technical University of Denmark, 2800 Kgs.\ Lyngby, Denmark}
\affiliation{Department of Micro- and Nanotechnology, Technical University of Denmark, 2800 Kgs.\ Lyngby, Denmark}
\author{Marin Solja\v{c}i\'{c}}
\affiliation{Department of Physics, Massachusetts Institute of Technology, Cambridge, Massachusetts, USA}
\author{N.~Asger Mortensen}
\affiliation{Center for Nanostructured Graphene, Technical University of Denmark, 2800 Kgs.\ Lyngby, Denmark}
\affiliation{Department of Photonics Engineering, Technical University of Denmark, 2800 Kgs.\ Lyngby, Denmark}



\begin{abstract}
The classical treatment of plasmonics is insufficient at the nanometer-scale due to quantum mechanical surface phenomena. Here, an extension to the classical paradigm is reported which rigorously remedies this deficiency through the incorporation of first-principles surface response functions -- the Feibelman $d$-parameters --  in general geometries. 
Several analytical results for the leading-order plasmonic quantum corrections are obtained in a first-principles setting; particularly, a clear separation of the roles of shape, scale, and material is established. The utility of the formalism is illustrated by the derivation of a modified sum-rule for complementary structures, a rigorous reformulation of Kreibig's phenomenological damping prescription, and an account of the small-scale resonance-shifting of simple and noble metal nanostructures. 
These insights open the technological design space and deepen our fundamental understanding of nanoplasmonics beyond the classical regime.
\end{abstract}

\maketitle
Classical treatments of plasmonics require specification of just two elements: geometry, involving shape and scale, and dielectric environment, supplied through local bulk dielectric functions. In the deep subwavelength regime, \ie in the nonretarded limit, even the element of scale is rendered superfluous by scale-invariant governing equations. As the geometric scale is reduced further, below $10-20\text{ nm}$ in metals, toward the intrinsic quantum mechanical length scales of the plasmon-supporting electron gas, the classical approach inevitably deteriorates, as established by numerous experiments~\cite{Genzel:1975,Ouyang:1992,Tiggesbaumker:1993,Reiners:1995,Charle:1998,Scholl:2012,Savage:2012,Raza:2013_Nanophotonics,Raza:2015c,Jin:2015}. The main shortcomings of the classical approach can be divided into three categories~\cite{Feibelman:1982}, resulting from the neglect of (i)~spill-out of the conduction electron's wave function beyond the material boundaries~\cite{Zhu:2016}, (ii)~nonlocality, \ie the momentum dependence of the bulk response functions~\cite{Raza:2015b}, and (iii)~incomplete accounting of internal electron dynamics, especially surface-enabled Landau damping~\cite{Khurgin:2014}. In the subnanometer domain additional shortcomings are expected to materialize, \eg due to size-quantization~\cite{Townsend:2011a,Townsend:2014} and the breakdown of jellium treatments~\cite{Varas:2016,Zhang:2014}.

Computationally, these shortcomings can be overcome by time-dependent density functional theory (TDDFT)~\cite{Marques:2006}, which, however, is limited to the study of few-atom clusters and systems of high spatial symmetry due to computational constraints. A sizable fraction of nanoplasmonic structures of interest~\cite{Schuller:2010,Novotny:2011,Giannini:2011,Teperik:2013} thus fall in a region which is simultaneously inaccessible to TDDFT and beyond the validity of classical plasmonics, roughly spanning characteristic geometric scales $L\sim 2-20\text{ nm}$. 
Here, we provide a simple and general answer to the central question raised by this dichotomy: namely, what are the leading-order nonclassical corrections to classical plasmonics at small $L$? 
We find that the three main shortcomings --~spill-out, nonlocality, and surface-enabled Landau damping~-- can be simultaneously overcome by extending the applicability of Feibelman's $d$-parameters~\cite{Feibelman:1982} to general geometries; an approach which is partly inspired by a recent computational development~\cite{Yan:2015}. Our simultaneous account of all three shortcomings is crucial; previous efforts to alleviate a solitary deficiency, \eg nonlocality within the hydrodynamic model~(HDM)~\cite{David:2011,Raza:2011,Ciraci:2012,Luo:2013}, are limited in scope and accuracy due to an arbitrary allocation of focus among nonclassical mechanisms of comparable magnitude.

The results presented here demonstrate that the leading-order spectral corrections to classical plasmonics appear as products of material-dependent surface response functions -- the Feibelman parameters $\dperp$ and $\dpar$ -- and a novel set of geometry-dependent perturbation factors, $\Lambda_{\sperp}^{\scriptscriptstyle(1)}$ and $\Lambda_{\spar}^{\scriptscriptstyle(1)}$, which exhibit a $1/L$ scale dependency. The resulting formalism, which amounts to a perturbation expansion of a generalized nonretarded boundary integral equation (nBIE), is simple and amenable to analytical treatments, yet rigorous and model-independent. The approach instates a natural partitioning of optical and electronic aspects, thereby indicating an advantageous division of labor in quantum nanoplasmonics between the condensed matter and optics communities.

\begin{figure}[!b]
	\centering
	\includegraphics[scale=1.085]{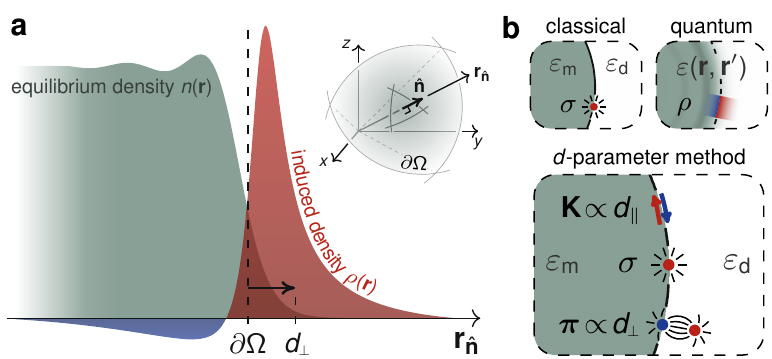}
	\caption{Surface features in quantum plasmonics: \textbf{a}, schematic of equilibrium and induced densities, $n(\rv)$ and $\rho(\rv)$, (distinct scales) plotted along a coordinate line, $\mathbf{r}_{\hat{\mathbf{n}}}$, normal to an $\hat{\mathbf{n}}$-oriented surface $\partial\Omega$ which delimits the ionic boundary of a metallic domain $\Omega$, see inset. Both $n(\rv)$ and $\rho(\rv)$ may extend beyond $\partial\Omega$; $\dperp$ is the centroid of $\rho(\rv)$.
		\textbf{b},~the leading-order differences between classical and quantum accounts of the plasmonic response of a surface may be bridged by introducing nonclassical contributions due to surface dipole and current densities, $\boldsymbol{\pi}(\rv)$ and $\mathbf{K}(\rv)$, normal and tangential to $\partial\Omega$ and proportional to the Feibelman parameters $\dperp$ and $\dpar$, respectively, which originate from a dipole expansion of $\rho(\rv)$.\label{fig:schematic}}
\end{figure}

\vskip 1.5em

\emph{Feibelman $d$-parameters---}
The classical local response (LR) description of light-scattering at an interface, say, a planar interface at $x=0$ separating metallic ($x<0$) and dielectric ($x>0$) regions with LR bulk dielectric functions $\varepsilon_{\text{m}}(\omega)$ and $\varepsilon_{\text{d}}(\omega)$, respectively, implies that the induced charge density, $\rho(\rv)$, is confined strictly to the interface such that $\rho(\rv)= \delta(x)\sigma(y,z)$.
The classical treatment consequently amounts to a monopole approximation of the nonsingular quantum mechanical $\rho(\rv)$, see Fig.~\ref{fig:schematic}. As demonstrated in Feibelman's seminal work on planar semi-infinite systems~\cite{Feibelman:1982}, the first-order extension of this zeroth-order multipole expansion naturally introduces two auxiliary quantities, $\dperp$ and $\dpar$, which parametrize the first moments of the induced charge and current density, $\mathbf{J}(\rv)$. A self-contained introduction to their properties is provided in the Supporting Material (SM)~\cite{Note0}. In brief, they represent model-dependent (\eg, TDDFT or HDM) surface-response functions,
and provide the leading-order corrections to classicality. Formally, for an external exciting potential $\phi^{\text{ext}}(\rv) = \e^{\iu ky+ kx}$ oscillating at frequency $\omega$, they follow directly from the induced dynamic quantities $\rho(\rv) = \rho(x)\e^{\iu ky}$ and $\mathbf{J}(\rv) = \mathbf{J}(x)\e^{\iu ky}$~\cite{Liebsch:1997}:
\begin{align}
\dperp=\frac{\int_{-\infty}^{\infty} x\rho(x)\dde{x}}{\int_{-\infty}^{\infty}\rho(x)\dde{x}},
\qquad
\dpar \equiv \frac{\int_{-\infty}^{\infty} x\tfrac{\partial}{\partial x} \mathrm{J}_{y}(x)\dde{x}}{\int_{-\infty}^{\infty} \tfrac{\partial}{\partial x} \mathrm{J}_{y}(x)\dde{x}}.
\end{align}
Both quantities define a characteristic length-scale of the dynamic problem: the centroid of induced charge ($\dperp$) and of the normal derivative of tangential current ($\dpar$)~\cite{Note1}. %
Notably, $\dpar$ vanishes for neutral strictly planar interfaces~\cite{Apell:1981,Liebsch:1997}, leaving $\dperp$ as the main quantity of interest for intrinsic quantum mechanical corrections; nevertheless, $\dpar$ is retained since it facilitates treatment of surface roughness~\cite{Apell:1984}, excess surface charge \eg due to adsorption, and semiclassical accounts of bound screening~\cite{Liebsch:1995}. Lastly, we note that the $d$-parameters are implicit functions of both $k$ and $\omega$; the $k$-dependence, however, is weak~\cite{Yan:2015}, and furthermore contributes only at second order in deviations from classicality, as observed first by \citet{Apell:1982a,Apell:1982b}. Crucially, this facilitates a mapping of local ($k\rightarrow 0$) $d$-parameters of \emph{planar} interfaces to general, curved geometries. This freedom of mapping is the central notion which allows the ensuing considerations~\cite{Note2}.

\vskip .25em
\emph{Governing equations---}
The classical nonretarded boundary integral equation (nBIE)~\cite{Kellogg:1967,Ouyang:1989,Abajo:1997} amounts to the solution of a scalar integral equation in an unknown surface charge density $\sigma(\rv,\omega)$ over a (possibly disconnected) surface domain $\rv\in\partial\Omega$, separating an interior metallic domain $\Omega$, with outward normal $\hat{\mathbf{n}}$, from an exterior dielectric domain. It constitutes a natural point of departure because it explicates the distinct and decoupled roles of material and shape as well as the scale-invariance of classical nonretarded treatments. The extension to account for surface contributions due to $\dperp$ and $\dpar$ follows by including two distinct polarizable boundary layers, see SM, one carrying a dipole density $\boldsymbol{\pi}(\rv,\omega) \equiv \dperp(\omega)\sigma(\rv,\omega)\hat{\mathbf{n}}$ and one carrying a surface current $\mathbf{K}(\rv,\omega) \equiv s(\omega)\mathbf{E}_{\spar}(\rv,\omega)$ proportional to the tangential electric field $\mathbf{E}_{\spar} \equiv (\mathbf{I}-\hat{\mathbf{n}}\hat{\mathbf{n}})\mathbf{E}$ and a surface conductivity $s(\omega) \equiv \iu\varepsilon_0(\varepsilon_{\text{d}}-\varepsilon_{\text{m}})\omega\dpar(\omega)$. The integral equation consistent with these additional terms is derived in the SM, and yields a generalized nBIE ($\omega$-dependence implicit)
\begin{subequations}\label{eqs:genbem}
\begin{align}
\Lambda \sigma(\rv) ={}& \mathcal{P}\!\int_{\partial\Omega}\!\!  \big[\nv\cdot  \boldsymbol{\nabla} g(\rv,\rv')\big]\sigma(\rv') \dde{^2\rv'} \nonumber  \\
&+\dperp \lim_{\delta\rightarrow 0^+}\! \int_{\partial\Omega}\!\! \big[\nv \cdot\boldsymbol{\nabla}\boldsymbol{\nabla}' g(\rv+\delta\nv,\rv')\cdot \nv'\big]\sigma(\rv')\dde{^2\rv'} \nonumber \\
&-\dpar\int_{\partial\Omega}\nabla_{\spar}^2g(\rv,\rv')\sigma(\rv')\dde{^2\rv'},
\label{eq:governing}
\end{align}
with scalar Coulomb interaction $g(\rv,\rv')\equiv 1/|\rv-\rv'|$, Cauchy principal value $\mathcal{P}$, surface Laplacian $\nabla_{\spar}^2$, and dimensionless eigenvalue $\Lambda$ parametrized by the frequency-dependent LR bulk dielectric functions of the constituent materials
\begin{equation}
\Lambda \equiv 2\pi\frac{\varepsilon_{\text{d}} + \varepsilon_{\text{m}}}{\varepsilon_{\text{d}} - \varepsilon_{\text{m}}}.
\end{equation}
\end{subequations}

Equation~\eqref{eq:governing} may equivalently be written in operator form as $\Lambda|\sigma\rangle = (\mathsf{K} + d_\alpha\mathsf{V}_{\alpha})|\sigma\rangle$, with operators $\mathsf{K}$ and $d_\alpha\mathsf{V}_{\alpha}$ (implicitly summed over $\alpha=\{\perp,\parallel\}$) acting on ket-states $\langle\rv|\sigma\rangle\equiv\sigma(\rv)$. The classical operator $\mathsf{K}$ is scale invariant, \cf its nondimensionalized form. Accordingly, the classical eigenproblem $\Lambda^{\scriptscriptstyle(0)}|\sigma^{\scriptscriptstyle(0)}\rangle = \mathsf{K}|\sigma^{\scriptscriptstyle(0)}\rangle$ is solely shape-dependent and its dimensionless eigenvalues $\Lambda^{\scriptscriptstyle (0)}$ constitute plasmonic shape factors. Conversely, the nonclassical operators $\mathsf{V}_{\alpha}$ exhibit an inverse scale dependency $\propto\! 1/L$, thereby introducing scale-invariance breaking of magnitude $d_\alpha/L$. Even so, for small but non-negligible breaking, the spectral properties remain expressible in terms of shape factors, as we demonstrate in the following.

\vskip .25em
\emph{Nonclassical geometry-dependent corrections---}
The eigensolutions $\{\Lambda_n^{\scriptscriptstyle(0)},|\sigma_n^{\scriptscriptstyle (1)}\rangle\}$ of $\mathsf{K}$, and their associated surface potentials $|\phi_n^{\scriptscriptstyle(0)}\rangle \equiv (4\pi\varepsilon_0)^{-1}\mathsf{g}|\sigma_n^{\scriptscriptstyle(0)}\rangle$ [with $\langle\rv|\mathsf{g}|\rv'\rangle \equiv g(\rv,\rv')$] form a biorthogonal basis over $\partial\Omega$ such that $\langle \phi_n^{\scriptscriptstyle (0)}|\sigma_{n'}^{\scriptscriptstyle(0)}\rangle\propto\delta_{nn'}$~\cite{Ouyang:1989}.
In seeking the leading order corrections to $\Lambda^{\scriptscriptstyle(0)}$ due to $d_\alpha \mathsf{V}_\alpha$, we may consequently apply perturbation theory around these classical eigensolutions. Specifically, writing the perturbed eigenvalue $\Lambda$ as (eigen-index $n$ implicit)
\begin{subequations}\label{eqs:perturbativelambda}
\begin{equation}\label{eq:corrected_lambda}
\Lambda = \Lambda^{\scriptscriptstyle(0)} +
\Lambda_{\alpha}^{\scriptscriptstyle(1)}d_\alpha +
\mathcal{O}(d_{\alpha}^2),
\end{equation}
introduces geometry-dependent perturbation factors $\Lambda_\alpha^{\scriptscriptstyle(1)} \equiv {\langle\phi^{\scriptscriptstyle(0)}| \mathsf{V}_\alpha |\sigma^{\scriptscriptstyle(0)}\rangle}/ {\langle\phi^{\scriptscriptstyle(0)}|\sigma^{\scriptscriptstyle(0)}\rangle}$ which simplify to (see SM)
\begin{align}\label{eq:perturbativelambda_perp}
\Lambda_{\sperp}^{\scriptscriptstyle(1)} &= \frac{(\Lambda^{\scriptscriptstyle (0)})^2-(2\pi)^2}{4\pi\varepsilon_0}\frac{\langle \sigma^{\scriptscriptstyle (0)}|\sigma^{\scriptscriptstyle (0)}\rangle}{\langle\phi^{\scriptscriptstyle (0)}|\sigma^{\scriptscriptstyle (0)}\rangle},\\ \label{eq:perturbativelambda_par}
\Lambda_{\spar}^{\scriptscriptstyle(1)} &= 4\pi\varepsilon_0\frac{\langle\boldsymbol{\nabla}_{\!\spar}\phi^{\scriptscriptstyle(0)}|\boldsymbol{\nabla}_{\!\spar}\phi^{\scriptscriptstyle(0)}\rangle}{\langle\phi^{\scriptscriptstyle(0)}|\sigma^{\scriptscriptstyle (0)}\rangle}.
\end{align}
\end{subequations}
We note the following features of these parameters:
(i) their unit is inverse length and they thus represent effective wave numbers, analogous to $k$ in the planar semi-infinite system;
(ii) nondimensionalization reveals a factorizable form $\Lambda_\alpha^{\scriptscriptstyle(1)} = \tilde{\Lambda}_\alpha^{\scriptscriptstyle(1)}/L$ in terms of a dimensionless shape factor $\tilde{\Lambda}_\alpha^{\scriptscriptstyle(1)}$ and a characteristic scale $1/L$; 
(iii) $\Lambda_{\sperp}^{\scriptscriptstyle(1)}<0$ and $\Lambda_{\spar}^{\scriptscriptstyle(1)}>0$, see SM; and
(iv) they are ratios of energies of the classical mode $|\sigma^{\scriptscriptstyle(0)}\rangle$, specifically, $\Lambda_{\sperp}^{\scriptscriptstyle(1)}$ and $\Lambda_{\spar}^{\scriptscriptstyle(1)}$ are proportional to the energy in the fictitious dipole- and current-layers, respectively, relative to the potential energy of classical resonance.

The perturbation result for $\Lambda$, Eqs.~\eqref{eqs:perturbativelambda}, allows a concomitant spectral statement. Specifically, for a classical eigenfrequency $\omega^{\scriptscriptstyle (0)} \equiv \omega(\Lambda^{\scriptscriptstyle (0)})$, the first-order spectral correction $\omega \equiv \omega^{\scriptscriptstyle (0)} + \omega^{\scriptscriptstyle(1)} + \mathcal{O}[(\omega-\omega^{\scriptscriptstyle(0)})^2]$ follows by expanding Eq.~\eqref{eq:corrected_lambda}~around~$\omega^{\scriptscriptstyle (0)}$:%
\begin{equation}\label{eq:spectralcorrection_1st}
\omega^{\scriptscriptstyle(1)} =
\frac{  \Lambda_{\alpha}^{\scriptscriptstyle(1)}d_{\alpha}^{\scriptscriptstyle(0)} }
	 {  \tfrac{\partial}{\partial\omega}\big( \Lambda -
	 	\Lambda_{\alpha}^{\scriptscriptstyle(1)}d_\alpha \big)^{\scriptscriptstyle(0)}  }
\simeq
\frac{  \Lambda_{\alpha}^{\scriptscriptstyle(1)}d_{\alpha}^{\scriptscriptstyle(0)} }
	{  \big(\tfrac{\partial}{\partial\omega}\Lambda\big)^{\scriptscriptstyle(0)}  },
\end{equation}
here the second approximate equality neglects the dispersion of $d_{\alpha}(\omega)$, i.e.\ a pole-approximation, and the superscript $(0)$ indicates evaluation at the classical frequency $\omega^{\scriptscriptstyle (0)}$, such that \eg $d_{\alpha}^{\scriptscriptstyle(0)} \equiv d_{\alpha}(\omega^{\scriptscriptstyle(0)})$.
The result is particularly elucidating for the lossless homogeneous electron gas (HEG) in vacuum [$\varepsilon_{\text{m}}(\omega) = 1-\omega_{\text{p}}^2/\omega^2$ and $\varepsilon_{\text{d}} = 1$], reducing there to $\omega^{\scriptscriptstyle(1)} = \tfrac{1}{8\pi}\Lambda_{\alpha}^{\scriptscriptstyle(1)}d_{\alpha}^{\scriptscriptstyle(0)} \omega_{\text{p}}^2/\omega^{\scriptscriptstyle(0)}$. Since $\Lambda_{\sperp}^{\scriptscriptstyle(1)}< 0$ this demonstrates that resonances redshift (blueshift) if $\dperp^{\scriptscriptstyle(0)}>0$ ($<0$), paralleling the results of the planar interface. Conversely, the sign of $\dpar^{\scriptscriptstyle(0)}$ indicates shifting in the opposite direction since $\Lambda_{\spar}^{\scriptscriptstyle(1)}> 0$.

\begin{table}[!tb]
	\def\tabsizeA{\small}
\def\tabsizeB{\footnotesize}
\def\tabsizeC{\ssmall}
\def\xw{1.2}
\def\h{.45}
\def\ch{.225}
\def\cw{1.1}
\def\r{.375}
\def\colw{1.3cm}
\tabcolsep=0.11cm

\begin{tabular}{>{\raggedleft\arraybackslash} m{1.2cm} 
				>{\centering\arraybackslash}  m{1.9cm}
				>{\centering\arraybackslash}  m{1.9cm}
				>{\centering\arraybackslash}  m{2.6cm}} 
\toprule
\tabsizeA Geometry &
\tabsizeA $\tfrac{1}{2\pi}\Lambda^{\scriptscriptstyle (0)}$ &
\tabsizeA $\tfrac{1}{2\pi}\Lambda_{\sperp}$ &
\tabsizeA $\tfrac{1}{2\pi}\Lambda_{\spar}$ \\
\midrule 
\begin{tikzpicture}
\fill [white] (-\xw/2,\h/1.15) rectangle (\xw/2,0);
\shade (-\xw/2,-\h/1.15) rectangle (\xw/2,0);
\draw (-\xw/2,0) -- (\xw/2,0);
\draw [dash pattern = on 1.25pt off 1.75pt] (-\xw/2,-\h/1.15) -- (\xw/2,-\h/1.15);
\draw (-\xw/2,0) node [below right,inner sep = 0.2pt] {\tabsizeC\textsf{half-space}};
\end{tikzpicture}  
& \tabsizeA $0$
& \tabsizeA $-k$
& \tabsizeA $k$\\
\begin{tikzpicture}
\fill [white] (-\xw/2,-\h/1.15) rectangle (\xw/2,\h/1.15);
\shade (-\xw/2,-\h/2.5) rectangle (\xw/2,\h/2.5);
\draw [mygray,thin] (\xw/3,\h/2.5) -- (\xw/3,.21*\h); 
\draw [mygray,thin] (\xw/3,-\h/2.5) -- (\xw/3,-.21*\h); 
\draw [mygray] (\xw/3,0) node {\ssmall $t$};
\draw (-\xw/2,-\h/2.5) -- (\xw/2,-\h/2.5);
\draw (-\xw/2,\h/2.5) -- (\xw/2,\h/2.5);
\draw (-\xw/2,\h/2.5) node [below right,inner sep = 0.2pt] {\tabsizeC\textsf{slab}};
\end{tikzpicture}
& \tabsizeA $\mp \e^{-k t}$
& \tabsizeA $-(1\mp \e^{-kt})k$
& \tabsizeA $(1\pm \e^{-kt})k$\\
\begin{tikzpicture}
\shade [top color=gray,bottom color=white] (-\xw/2,-\h/1.15) rectangle (\xw/2,-\h/2.5);
\shade [top color=white,bottom color=gray] (-\xw/2,\h/1.15) rectangle (\xw/2,\h/2.5);
\draw [mygray,thin] (\xw/3,\h/2.5) -- (\xw/3,.21*\h); 
\draw [mygray,thin] (\xw/3,-\h/2.5) -- (\xw/3,-.21*\h); 
\draw [mygray] (\xw/3,0) node {\ssmall $t$};
\draw (-\xw/2,-\h/2.5) -- (\xw/2,-\h/2.5);
\draw (-\xw/2,\h/2.5) -- (\xw/2,\h/2.5);
\draw [dash pattern = on 1.25pt off 1.75pt] (-\xw/2,\h/1.15) -- (\xw/2,\h/1.15);
\draw [dash pattern = on 1.25pt off 1.75pt] (-\xw/2,-\h/1.15) -- (\xw/2,-\h/1.15);
\draw (-\xw/2,\h/2.5) node [above right,inner sep = 0.2pt] {\tabsizeC\textsf{gap}};
\end{tikzpicture}
& \tabsizeA $\pm \e^{-k t}$
& \tabsizeA $-(1\mp \e^{-kt})k$
& \tabsizeA $(1\pm \e^{-kt})k$\\
\begin{tikzpicture}
\draw [mygray,thin,densely dotted] ({.05*\r},\r) -- ({1.5*\r},\r);
\draw [mygray,thin,densely dotted] ({.05*\r},-\r) -- ({1.5*\r},-\r);
\draw [mygray,thin] ({1.5*\r},\r) -- ({1.5*\r},-\r) node [midway,fill=white,inner sep=1pt] {\ssmall $2R$};

\shadedraw (0,0) circle (\r cm);
\draw [densely dashed] (\r,0) arc (0:-115:{\r} and \r/2);
\draw [densely dashed] (0,\r) arc (90:-90:{\r/5} and \r);
\def\myshift#1{\raisebox{-.225ex}}
\path [postaction={decorate,decoration={text along path,text align=center,text={|\tabsizeC\sffamily\myshift|sphere}}}] ({\r*cos(230)*.675},{\r*sin(230)*.675}) arc (230:80:{\r*.675});
\end{tikzpicture}    
& \tabsizeA $\displaystyle-\frac{1}{2l+1}$
& \tabsizeA $\displaystyle-\frac{2 l(l+1)}{(2l+1)R}$
& \tabsizeA $\displaystyle \frac{2 l(l+1)}{(2l+1)R}$\\
\begin{tikzpicture}
\shade (\cw/2,\ch) arc (90:-90:{\ch/3} and \ch) -- (-\cw/2,-\ch) arc (-90:90:{\ch/3} and \ch) -- (\cw/2,\ch) -- cycle;
\draw [dash pattern = on 1.25pt off 1.75pt] (\cw/2,\ch) arc (90:-90:{\ch/3} and \ch) ;
\shade [left color=gray,right color=white,fill opacity = .5] (-\cw/2,\ch) arc (90:450:{\ch/3} and \ch) ;
\draw [dash pattern = on 1.25pt off 1.75pt] (-\cw/2,\ch) arc (90:360:{\ch/3} and \ch) ;
\draw (-\cw/2,\ch) -- (\cw/2,\ch);
\draw (-\cw/2,-\ch) -- (\cw/2,-\ch);
\draw [densely dashed] (\cw/4,\ch) arc (90:-90:{\ch/3} and \ch) ;
\draw [densely dashed] (-\cw/4,-\ch) arc (-90:20:{\ch/3} and \ch);
\def\cmx{\cw/40}
\def\rotlab{-17}
\coordinate (A) at ({-\cmx+cos(\rotlab)*.2},{\ch + sin(\rotlab)*.2}); 
\coordinate (B) at ({-\cmx+cos(\rotlab)*.2},{-\ch + sin(\rotlab)*.2}); 
\draw [mygray,thin,densely dotted] (-\cmx,\ch) --  (A);
\draw [mygray,thin,densely dotted] (-\cmx,-\ch) -- (B);
\draw [mygray,thin] (A) -- ({-\cmx+cos(\rotlab)*.2},{.5*\ch + sin(\rotlab)*.2});
\draw [mygray,thin] (B) -- ({-\cmx+cos(\rotlab)*.2},{-.5*\ch + sin(\rotlab)*.2});
\draw [mygray] ({-\cmx+cos(-15)*.2},{sin(\rotlab)*.2}) node {\ssmall $2R$};
\draw (-\cw/2-\cw/22.5,\ch) node [below right,inner sep = 0.2pt] {\tabsizeC\textsf{cylinder}};
\end{tikzpicture}  
& \tabsizeB $\tilde{k}\big[K_m(\tilde{k})I_{m}(\tilde{k})\big]'$ 
& \tabsizeB $\displaystyle\frac{\big(\tfrac{1}{2\pi}\Lambda^{\scriptscriptstyle(0)}\big)^2 - 1}{2 K_m(\tilde{k})I_m(\tilde{k}) R}$ 
& \tabsizeB $\displaystyle 2K_m(\tilde{k})I_m(\tilde{k})\frac{m^2+\tilde{k}^2}{R}$\\
\bottomrule
\end{tabular}
	\caption{Exact analytical eigenvalues $\Lambda\equiv\Lambda^{\scriptscriptstyle(0)} + \dperp\Lambda_{\sperp}+\dpar\Lambda_{\spar}$ of Eq.~\eqref{eq:governing} valid to all orders in $d_{\alpha}$. The metallic geometries (and associated geometric length scales) are indicated schematically in gray; the relevant eigen-indices are, from top to bottom, wave number $k$, symmetric (upper sign) and antisymmetric (lower sign) charge density parity, polar angular momentum $l$, azimuthal angular momentum $m$, and dimensionless axial wave number $\tilde{k}\equiv kR$. $K_m$ and $I_m$ denote modified Bessel functions.\label{tab:analyticalresults}}
\end{table}

In systems of sufficiently high symmetry, the perturbative results, \ie Eq.~\eqref{eqs:perturbativelambda}, coincide with exact solutions of Eq.~\eqref{eq:governing} since first- and higher-order corrections to $|\sigma\rangle = |\sigma^{\scriptscriptstyle(0)}\rangle + |\sigma^{\scriptscriptstyle(1)}\rangle + \ldots$ vanish by symmetry constraints. Table~\ref{tab:analyticalresults} lists exact analytical results for a number of such sufficiently symmetric systems, derived using suitable modal expansions of the Coulomb interaction, see SM. The results for the half-space and sphere reproduce the special cases previously obtained by \citet{Feibelman:1982} and \citet{Apell:1982a,Apell:1982b}, respectively. The generality of the present approach additionally allows the derivation of new analytical results, here exemplified for the cylinder, slab, and gap geometries.
The utility and universality of the present approach is further illustrated by the fact that Eqs.~\eqref{eqs:perturbativelambda} and~\eqref{eq:spectralcorrection_1st}, and Table~\ref{tab:analyticalresults} in particular, readily reproduce all known first-order HDM results~\cite{Yan:2013,Schnitzer:2016} when the HDM approximation of the $d$-parameters is employed~\cite{Note3}, \ie when $\dpar^{\text{\textsc{hdm}}}=0$ and $\dperp^{\text{\textsc{hdm}}}(\omega) = -\beta/(\omega_{\text{p}}^2-\omega^2)^{1/2}$ with $\beta^2\equiv \tfrac{3}{5}\vf^2$~\cite{Feibelman:1982}.

\begin{figure}[!tb]
	\centering
	\includegraphics{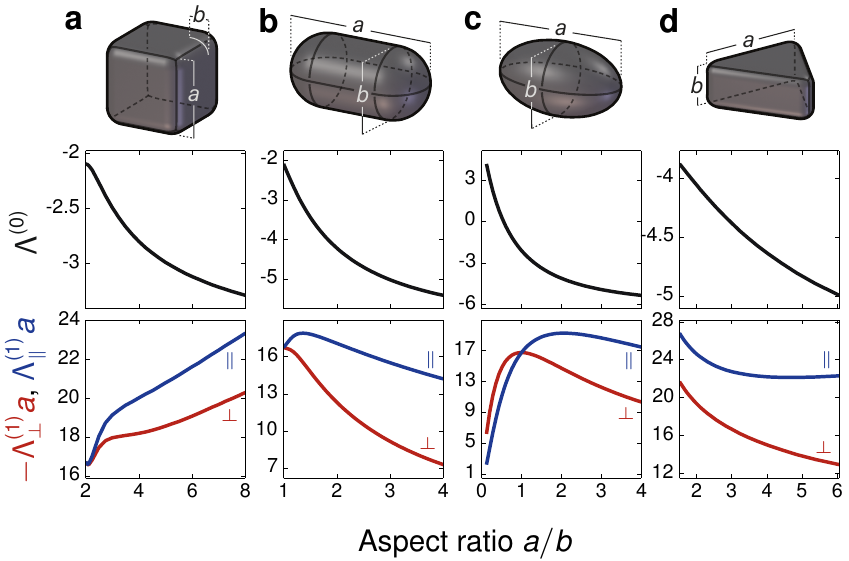}
	\caption{Aspect ratio dependence, $a/b$, of the shape factors, $\Lambda^{\scriptscriptstyle(0)}$, $-\Lambda^{\scriptscriptstyle(1)}_\sperp a$, and $\Lambda^{\scriptscriptstyle(1)}_\spar a$, with the latter two normalized by the length $L=a$, for four canonical geometries. The values correspond to the dipole resonances of \textbf{a}, cubes of side $a$ and edge- and corner-rounding $2b$, \textbf{b}, cylindrical pills of length $a$, diameter $b$, and butt-rounding $b$, \textbf{c}, spheroids with principal axis $a$ and $b$, and \textbf{d}, equilateral triangles of height $b$, side $a$, and edge- and corner-rounding $\approx 0.165a$. Mode polarization is along $a$ in \textbf{a}--\textbf{c} and along the triangle altitude in  \textbf{d}. Rounding is intramural and of cylindrical and spherical kind with inscribed diameters equaling the specified rounding value.\label{fig:opticalgeometry}}
\end{figure}

In less symmetric geometries analytical solutions cannot generally be obtained. Regardless, the classical nBIE operator $\mathsf{K}$ can be discretized by the boundary element method~\cite{Abajo:1997} allowing the numerical calculation of the nonclassical shape factors $\tilde{\Lambda}_{\alpha}^{\scriptscriptstyle(1)}=\Lambda_{\alpha}^{\scriptscriptstyle(1)}L$ via Eqs.~\eqref{eqs:perturbativelambda}. Figure~\ref{fig:opticalgeometry} presents the results of such a calculation, here for the dipolar modes of experimentally relevant geometries over a range of aspect ratios $a/b$, specifically for cubes, pills, spheroids, and triangles. The former three reduce to spheres at aspect ratios $a/b = 2$, $1$, and $1$, respectively.
Interestingly, though the $a/b$ dependence of the classical dipole eigenvalue $\Lambda^{\scriptscriptstyle(0)}$ is qualitatively similar across the considered shapes, \eg monotonically decreasing with $a/b$, the corresponding dependence of $\Lambda_\alpha^{\scriptscriptstyle(1)}$ is markedly dissimilar for distinct shapes.
In this sense, nonclassicality constitutes a stronger probe of local geometric features than its underlying classical correspondent.

\vskip .25em

\emph{Breaking of classical complementarity---}
The classical nBIE naturally leads to a nonretarded spectral sum-rule for the resonances of complementary geometries (\ie, of interchanged material regions)~\cite{Apell:1996}. Concretely, the equimodal (\ie, of identical modal pattern) eigenvalues of a region $\Omega$ and its complement $\Omega^{\scriptscriptstyle\complement}\equiv\mathbb{R}^3\backslash\Omega$, denoted $\Lambda^{\scriptscriptstyle (0)\scriptsize}$ and $\Lambda^{\scriptscriptstyle (0)\scriptsize,\scriptscriptstyle\complement}$, respectively, are interrelated by $\Lambda^{\scriptscriptstyle (0)} = - \Lambda^{\scriptscriptstyle (0)\scriptsize,\scriptscriptstyle\complement}$, since $\Omega$ and $\Omega^{\scriptscriptstyle\complement}$ are distinguished in the nBIE only by the sign of the surface normal $\hat{\mathbf{n}}$~\cite{Apell:1996}. This is the classical statement of complementarity in the small-scale limit. The present extension of the nBIE allows a refinement of this statement; specifically, it follows from the absence of an $\hat{\mathbf{n}}$-dependence in Eqs.~\eqref{eqs:perturbativelambda} that $\Lambda^{\scriptscriptstyle (1)}_{\alpha} = \Lambda^{\scriptscriptstyle (1)\scriptstyle,\scriptscriptstyle\complement}_{\alpha}$ (this fact is exemplified, \eg, by the slab and gap results of Table~\ref{tab:analyticalresults}). Consequently, classical complementarity is broken in the sense
\begin{equation}\label{eq:plasmondecay}
\Lambda + \Lambda^{\scriptscriptstyle\complement}  =
\Lambda^{\scriptscriptstyle (1)}_{\alpha} \big(d_{\alpha}+d_{\alpha}^{\scriptscriptstyle\complement}\big) +
\mathcal{O}(d_{\alpha}^2),
\end{equation}
with $d_{\alpha}^{(\scriptscriptstyle\complement\scriptstyle)}$ evaluated at  $\omega^{(\scriptscriptstyle\complement\scriptstyle)}$.
For the HEG in vacuum, this entails a modified sum-rule $\omega^2+(\omega^{\scriptscriptstyle\complement})^2 \simeq \omega_{\text{p}}^2 [ 1 +  \tfrac{1}{2}\Lambda^{\scriptscriptstyle (1)}_{\alpha} (d_\alpha + d_\alpha^{\scriptscriptstyle\complement})]$.
This new finding establishes that classical complementarity is generically broken, even in the small-scale limit; it is attained only approximately in an intermediate domain bounded by large- and small-scale breakings due to retardation ($\appropto L$) and nonclassical surface effects ($\appropto 1/L$).
A prior HDM study of the slab-gap system constitutes a special case of this result~\cite{Raza:2013_PhysRevB}.

\begin{figure}[!b]
	\centering
	\includegraphics{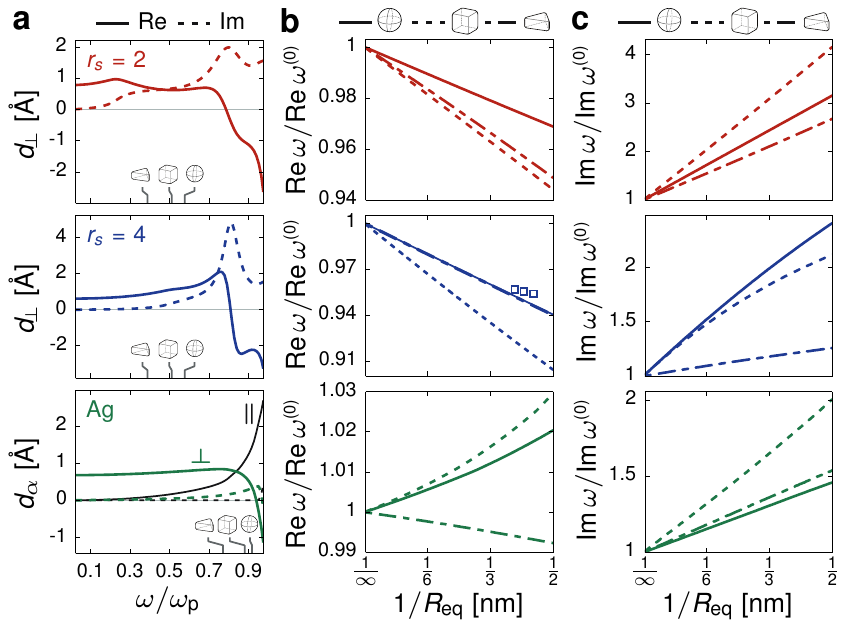}
	\caption{The complementary roles of material, shape, and scale for the spectral properties of plasmonic nanostructures: \textbf{a},~the spectral dependence of the $\dperp$ for HEGs ($r_s = 2$ and $4$) and Ag ($\dpar$ in black). Ag is normalized by its screened plasma frequency $\hbar\omega_{\text{p}} \equiv 3.81\text{ eV}$.
		\textbf{b} and \textbf{c},~the real and imaginary part of the resonance dispersion with inverse scale, indicated by equivalent spherical radius $R_{\text{eq}} \equiv (\tfrac{3}{4\pi}\times\text{volume})^{1/3}$, for spheres, cubes ($a/b = 5$), and triangles ($a/b = 3$) for the materials considered in \textbf{a} with matching color-code (HEGs; Drude decay $\gamma^{\scriptscriptstyle(0)} = \omega_{\text{p}}/50$).
		The classical resonance position $\text{Re}\,\omega^{\scriptscriptstyle(0)}$ of the geometries are indicated in \textbf{a} by line-connected symbols. TDDFT calculations by \citet{Weick:2006} of jellium Na nanospheres are shown by  square markers in \textbf{b} for comparison with the case $r_{s} = 4$.%
		\label{fig:explicitcalculation}%
	}
\end{figure}

\vskip .25em

\emph{Surface-enhanced plasmon decay---}
Finally, we discuss the size-dependent decay of plasmons.
Equation \eqref{eq:spectralcorrection_1st} directly facilitates a rigorous treatment of this aspect; in particular, splitting the imaginary part of a resonance frequency $\mathrm{Im}\,\omega \equiv -\tfrac{1}{2}(\gamma^{\scriptscriptstyle(0)} + \gamma^{\scriptscriptstyle(1)})$ into a classical part $\gamma^{\scriptscriptstyle(0)}$ due to bulk absorption and a nonclassical part $\gamma^{\scriptscriptstyle(1)}$ due to surface-enabled absorption, we find (assuming $\mathrm{Re}\,\omega\gg\gamma^{\scriptscriptstyle(0,1)}$)%
\begin{align}
\gamma^{\scriptscriptstyle(1)}
&\simeq -2\frac{\Lambda_\alpha^{\scriptscriptstyle(1)}\mathrm{Im}\,d_\alpha^{\scriptscriptstyle(0)}}{\mathrm{Re}\,\big(\tfrac{\partial}{\partial\omega}\Lambda\big)^{\scriptscriptstyle(0)}}
\simeq -\frac{1}{4\pi}\frac{\omega_{\mathrm{p}}^2}{\mathrm{Re}\,\omega^{\scriptscriptstyle(0)}}\Lambda_\alpha^{\scriptscriptstyle(1)}\mathrm{Im}\,d_\alpha^{\scriptscriptstyle(0)},
\end{align}%
specializing in the last equality to the HEG in a vacuum. This result generalizes the well-known phenomenological Kreibig approach often adopted in nanospheres which takes $\gamma^{\scriptscriptstyle(1)} \simeq \vf/R$~\cite{Kreibig:1969,Kreibig:1985} extending its applicability to arbitrary geometries~\cite{Note4}.
Similarly, it provides a first-principles alternative to the recently proposed diffusion-HDM `GNOR'~\cite{Mortensen:2014}.

These considerations are further expounded in Fig.~\ref{fig:explicitcalculation} for HEGs of Wigner--Seitz radius $r_s = 2$ and $4$ (qualitatively representative of Al and Na, respectively) and Ag. Figure~\ref{fig:explicitcalculation}a depicts TDDFT calculations of $\dperp$ (using the Gunnarsson--Lundqvist exchange-correlation potential~\cite{Gunnarsson:1976}). 
For Ag,  the 5s orbitals are treated at the TDDFT level, while d-band screening is treated semiclassically through Liebsch' method~\cite{Liebsch:1993}, see SM; this method necessitates inclusion of nonzero $\dpar$-values~\cite{Liebsch:1995}. 
The impact of these $d$-parameters on the spectral size-dispersion of plasmons is explored in Fig.~\ref{fig:explicitcalculation}b-c for a sphere, cube, and triangle, obtained by numerical solution of Eq.~\eqref{eq:corrected_lambda} with shape-factors from Table~\ref{tab:analyticalresults} and Fig.~\ref{fig:opticalgeometry}. 
Bulk dielectric properties of Ag is taken from measured data~\cite{Johnson:1972}. 
The considered HEGs exhibit a redshifting $\text{Re}\,\omega$ since $\dperp(\omega^{\scriptscriptstyle(0)})>0$; conversely, the interplay between $\dperp$ and $\dpar$ manifests itself as a blueshift for the Ag sphere and cube. 
Surprisingly, the Ag triangle redshifts revealing that this key characteristic may depend on geometric shape in addition to material. 
The inverse scale proportionality $\omega/\omega^{\scriptscriptstyle(0)}-1\appropto 1/L$ exemplified by Eq.~\eqref{eq:plasmondecay} is clearly displayed for both $\text{Re}\,\omega$ and $\text{Im}\,\omega$, being only slightly modified at smallest considered scale due to spectral dispersion of $d_{\alpha}$. 
The $r_s = 4$ nanosphere is compared with TDDFT calculations of $\text{Re}\,\omega$ by \citet{Weick:2006}; excellent agreement is observed, an observation that endures even at smaller radii (see SM), underlining the accuracy of the present approach.

\vskip .25em

\emph{Conclusions and outlook---}
The results presented in here demonstrate that the complicated and rich interplay between scale, shape, and material in quantum nanoplasmonics may be understood quantitatively through just five parameters: $L$, $\tilde{\Lambda}^{\scriptscriptstyle(1)}_{\alpha}$, and $d_{\alpha}$. These parameters are the natural nonclassical extensions that complement the bulk dielectric functions and modal shape factor, $\varepsilon_{\text{m}}$,  $\varepsilon_{\text{d}}$, and $\Lambda^{\scriptscriptstyle(0)}$, of classical plasmonics. They originate physically from dynamic surface dipole- and current-densities, $\boldsymbol{\pi}(\rv)$ and $\mathbf{K}(\rv)$, proportional to the Feibelman $d$-parameters, see Fig.~\ref{fig:schematic}b. Together, they provide a general and first-principles approach, which transparently and accurately separates the distinct roles of shape, scale, and material down to the nanometer scale. 

Several exciting aspects remain unexplored: for instance, the retarded generalization of this approach follows by including the same boundary terms $\boldsymbol{\pi}(\rv)$ and $\mathbf{K}(\rv)$,
allowing immediate incorporation \eg in retarded BEMs.
Another contiguous application lies with coupled nanostructures, with implications \eg for plasmon rulers~\cite{Teperik:2013}.
Nonclassical modifications to scattering properties~\cite{Note5} and their concomitant impact on classical sum-rules and scattering limits~\cite{Miller:2014} poses a separate open problem. 
Moreover, while the focus here has rested on the perturbative impact of nonclassicality, additional features without a classical equivalent are contained in the framework, such as the Bennett mode~\cite{Bennett:1970} corresponding to poles of $\dperp$~\cite{Tsuei:1991}.
Finally, the approach extends to several novel plasmonic platforms, such as highly doped semiconductors~\cite{Schimpf:2014} -- it may translate to 2D plasmonics as well, \eg enabling analytical insight in the plasmonic properties of zigzag- vs.\ armchair-terminated graphene nanostructures~\cite{Christensen:2014prb} through analogous nonclassical \emph{edge} densities.

In conclusion, we hope these results will renew interest in the Feibelman $d$-parameters as a general tool and fundamental platform in the field of quantum nanoplasmonics.

\vskip 1.5em
\begin{acknowledgments}
T.C.\ thanks Dafei Jin and Martijn Wubs for valuable discussions and acknowledges support from the Villum Foundation.
W.Y.\ acknowledges support from the Lundbeck Foundation, grant no.~70802.
The Center for Nanostructured Graphene is sponsored by the Danish National Research Foundation, Project DNRF103.
The work was also supported by the Danish Council for Independent Research--Natural Sciences, Project~1323-00087.
M.S.\ was funded in part (analysis and reading of the manuscript) by S3TEC an Energy Frontier Research Center funded by the U.S.\ Department of Energy, Office of Science, Office of Basic Energy Sciences under Award Number DE-SC0001299/DE-FG02-09ER46577.
\end{acknowledgments}


%

\end{document}